\documentclass[conference]{IEEEtran}
\makeatletter
\def\ps@headings{%
\def\@oddhead{\mbox{}\scriptsize\rightmark \hfil \thepage}%
\def\@evenhead{\scriptsize\thepage \hfil \leftmark\mbox{}}%
\def\@oddfoot{}%
\def\@evenfoot{}}
\makeatother
\usepackage{graphicx}
\usepackage{caption}
\usepackage{hyperref}

\newcommand{\descr}{\medskip\noindent\textbf}
\renewcommand{\paragraph}{\descr}

\begin{document}

\title{Optimizing Bi-directional Low-Latency Communication in Named Data Networking}

\author{\IEEEauthorblockN{Mishari Almishari \hspace{0.45cm} Paolo Gasti \hspace{0.45cm} Naveen Nathan \hspace{0.45cm} Gene Tsudik}
\IEEEauthorblockA{University of California, Irvine\vspace{-0.05cm}\\
{\tt\footnotesize \{malmisha,pgasti,nnathan,gts\}@uci.edu}}
}

\maketitle

\begin{abstract}
Content-Centric Networking (CCN) is a concept being considered as a potential
future alternative to, or replacement for, today's Internet IP-style packet-switched
host-centric networking. A key factor making CCN attractive is its focus on content distribution,
which dominates current Internet traffic and which is arguably not well-served by IP.
Named Data Networking (NDN) is a prominent example of CCN. It is also one of several
on-going research efforts aiming to design and develop a full-blown candidate future Internet
architecture. Although NDN's primary motivation is content distribution, it is envisioned to
support other types of traffic, such as conferencing (audio, video) as well as more historical
applications, such as remote login. 

However, it is unclear how suitable NDN is for applications that are not obviously
content-centric. We believe that such applications are not going away any time soon.
In this paper, we explore NDN in the context of a class of applications
that involve low-latency bi-directional (point-to-point) communication.
Specifically, we propose a few architectural amendments to NDN that provide 
significantly better throughput and lower latency for this class of applications.
The proposed approach is validated via experiments.
\end{abstract}

\section{Introduction}
Today's Internet serves as an effective platform for a multitude of applications, including WWW, Email, 
P2P and VoIP. Its main architectural pillar is IP~\cite{rfc791}, which follows the host-centric 
packet-switched communication paradigm where each host is referred via one or more 
interface addresses and communication is performed via IP packets. 
Although this model has proven to be better, and lasted longer, than originally envisioned, 
it is starting to fray. The Internet is being used to distribute greatly
increasing amounts of digital content. This unprecedented and long-lasting growth spurt
is due to the proliferation and popularity of multimedia content, 
social networks as well as increasing amounts of user-generated
content. The resulting fundamental change in the nature of Internet traffic
has exposed limitations of the current architecture. To this end, 
some projects aiming to design candidate next-generation Internet
architectures started within the last several years.

Named-Data Networking (NDN) \cite{NDN} is one such effort that
exemplifies the Content-Centric Networking (CCN) approach
\cite{gritter2001architecture,Jacobson2009,koponen2007data}. 
NDN explicitly names content instead of physical locations, such as hosts or network interfaces. 
Instead of a conversation-style semantics of IP where hosts directly address each other,
NDN applications request content via a meaningful human-readable name; the network is in 
charge of locating and returning the closest copy of requested content. 
(See Section \ref{sec:ndn-overview} for more NDN details.)
NDN also stipulates that each piece of named content must be digitally signed by its producer. 
This allows decoupling of trust in content from trust in entities that might
store and/or disseminate that content. 

NDN is primarily oriented towards efficient large-scale content distribution. 
However, in order to become a viable replacement for IP, it must also support other
types of Internet traffic. In other words, overall practicality of NDN depends, among
other things, on how it performs outside its {\em forte} of content distribution. 

This paper focuses on one specific type of Internet traffic that exhibits characteristics
very different from content distribution. It corresponds to  a class of applications
that involve low-latency bi-directional communication, such as audio and/or video 
conferencing as well as more traditional ones, such as remote login.
While we do not claim that this application class is not accommodated, or is poorly
supported by NDN, we believe that NDN is simply not attuned to these needs.
This prompts us to explore add-on techniques that might offer better performance.
As we show below, simple amendments that retain all basic NDN features (and do
not affect content distribution-type traffic) result in markedly improved end-to-end
throughput and bandwidth utilization. This assertion is supported by experiments. 

\paragraph{Organization.}
After a brief overview of NDN in the next section, we provide some motivation
for optimizing bi-directional communication in NDN, in Section~\ref{sec:motivation}. 
Then, in Section~\ref{sec:design}, we describe an interest-piggyback scheme that, 
while leaving key NDN features intact, offers markedly better performance for low-latency
point-to-point bi-directional communication.  We then discuss, in 
Section~\ref{sec:piggyback-implementation}, modifications to the NDN prototype 
to support our design. Section~\ref{sec:experiment} reports on experimental results 
that confirm claimed performance gains. 
We overview related work in Section~\ref{sec:related-work} and conclude in 
Section~\ref{sec:conclusion}.

\section{NDN Overview}
\label{sec:ndn-overview}
NDN supports two types of messages: {\em interests} and {\em content packets}~\cite{ccnx-protocol}. A content packet contains 
a human-readable name, actual data (content), and a digital signature computed by the content producer
over the packet. Names are hierarchically structured, e.g. \verb|/ndn/usa/cnn/frontpage/news| 
where ``\verb|/|'' is the boundary between name components. An interest packet contains the name of 
the content requested or prefix of such a name, e.g. \verb|/ndn/usa/cnn/| is a prefix of 
\verb|/ndn/usa/cnn/frontpage/news|. In case of multiple content under a given name prefix, 
optional control information can be carried within the interest to restrict the content returned. Content
signatures provide data origin authenticity, however trust management between a key
and a name prefix is the responsibility of the application.

All NDN communication is receiver-driven: a consumer initiates communication by sending 
an interest for a specific content. NDN routers forward this interest towards the content producer 
responsible for the requested name, using name prefixes (instead of today's IP prefixes) for
routing. {\em Forwarding Information Base} (FIB) is a lookup table used to determine interfaces 
for forwarding incoming interests, and contains [{\em name\_prefix}, {\em interface}] entries. 
Multiple entries with  the same {\em name\_prefix} are allowed, supporting multiple paths 
under which a given {\em name\_prefix} namespace is reachable. Akin to an IP forwarding table, 
FIB can be populated either by a routing protocol or manually.

Each NDN router maintains a Pending Interest Table (PIT), which is a lookup table containing 
outstanding [{\em interest}, {\em arrival-interfaces}] entries. The first component of a PIT
entry reflects the name of requested content, and the second -- a set
of interfaces via which interests for this content arrived. 

When an NDN router receives an interest, it first looks up its PIT to determine
whether an interest for the same named content is currently outstanding. There
are three possible outcomes: 

\begin{enumerate}
\item If the same name is already in the router's 
PIT and the arrival interface of the present interest is already in the set of 
{\em arrival-interfaces}  of the corresponding PIT entry, the interest is discarded. 

\item  If a PIT entry for the same name exists, yet the arrival interface is new, the
router updates the PIT entry by adding a new interface to the set; the interest is not forwarded further.

\item Otherwise, the router creates a new PIT entry and forwards the present
interest using its FIB.
\end{enumerate}

Upon receipt of the interest, the producer injects content into the network, thus 
{\em satisfying} the interest. The requested content is then forwarded towards the 
consumer, traversing -- in reverse -- the path of the corresponding interest. Each router 
on the path flushes state (deletes the PIT entry) containing the satisfied interest and 
forwards the content out on all arrival interfaces of the associated interest.
In addition, each router caches a copy of forwarded content in its 
local Content Store (CS). Unlike their IP counterparts, NDN routers can 
forward interests out on multiple interfaces in order to maximize the chances 
of quickly retrieving requested content. 

The above description of interest forwarding only applies to content that has not been 
recently requested, i.e., not present in CS-s of intervening routers. Whereas, 
a router that receives an interest for already-cached content does not forward the interest 
further; it simply returns cached content and retains no state about the interest.

Not all interests result in content being returned. If an interest wends its way
through the network and encounters either a router that cannot forward it further
or a content producer that has no such content, no error packets are
generated. PIT entries in intervening routers are simply to expire if no content can be retrieved.
The consumer (who also maintains its local PIT) can choose to regenerate the same
interest after a timeout.

\section{Motivation}
\label{sec:motivation}
As follows from the above discussion, the NDN architecture is primarily geared to applications 
that disseminate content. Routers directly assist content distribution by, whenever possible, 
satisfying consumer interests with cached content. This is very different from IP, since 
NDN decouples the flow of content  from the notion of content location. 

In this paper, we focus on bi-directional conversation-style applications over NDN.
Representative applications of this class are: audio/video conferencing, interactive chat, and 
remote login. 

\begin{figure}[htb]
\vspace{0.15cm}
\centering
\includegraphics[width=3.3in]{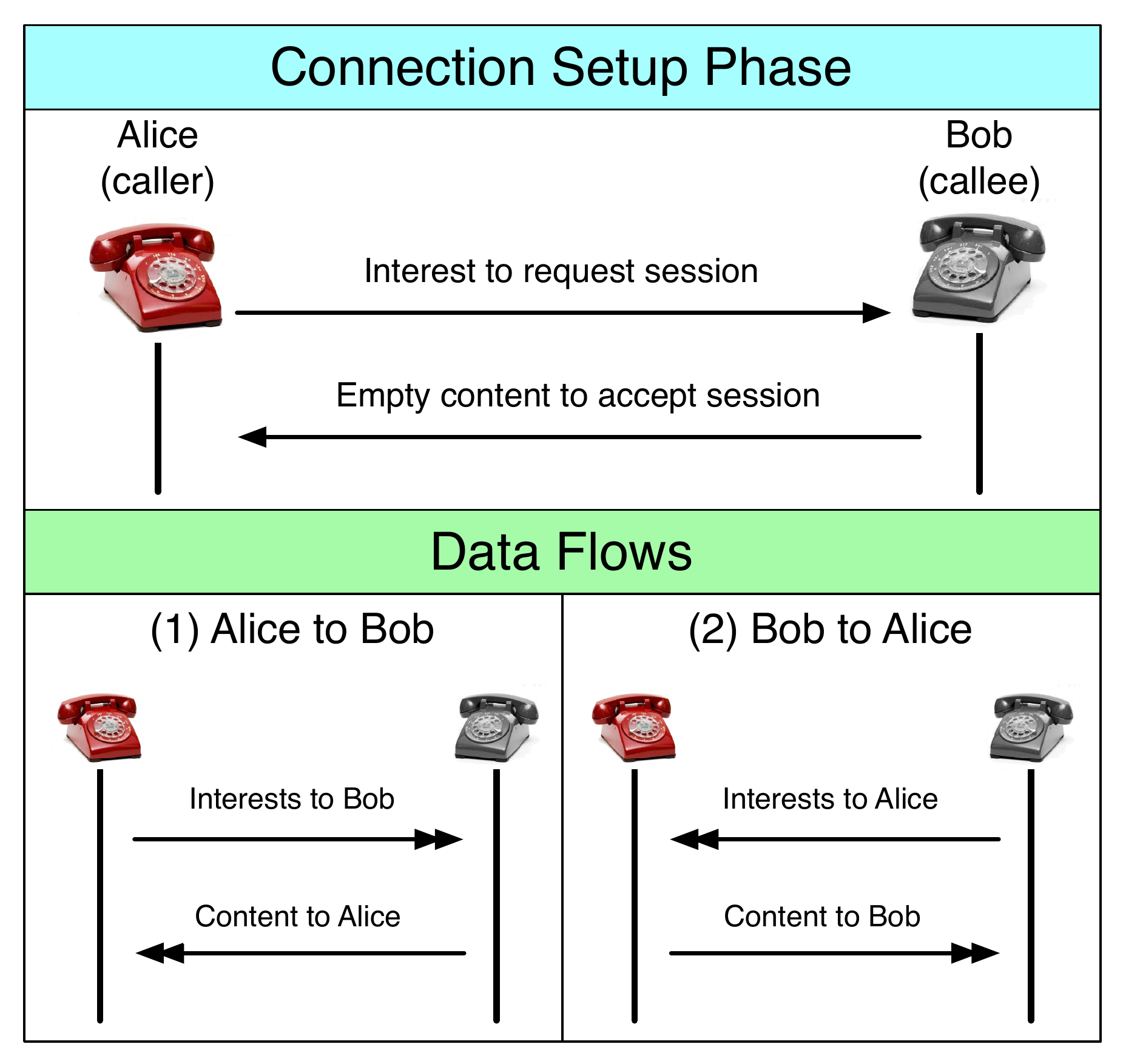}
\caption{Bi-directional communication between Alice and Bob.}
\label{fig:alice-bob-bidirectional}
\end{figure}

To support this communication paradigm in NDN, each end-point 
(e.g., Alice and Bob) must register its own namespace and, during the conversation, play the 
role of both producer and consumer of content.
Before describing the session establishment and data exchange protocol, we refer the reader to Figure \ref{fig:alice-bob-bidirectional}
for a high-level overview. 
To initiate a session, Alice issues an interest in Bob's namespace, embedding 
her own namespace prefix in the name (as a suffix). Bob receives the interest, 
parses the name which indicates that Alice is requesting a session. 
Bob then responds with a content acknowledging 
his agreement to establish a session. Thereafter, data flows in both directions:  
Alice and Bob exchange interests for each other's namespaces and generate content 
accordingly. Such a session can be viewed as two flows: Alice and Bob each request
(and receive) the other party's content via interests. NDN routers that forward interests and content 
between Alice and Bob are oblivious to the conversation. The two flows 
(Alice$\rightarrow$Bob and Bob$\rightarrow$Alice) might even wind up 
using asymmetric paths. This type of communication does not get the main 
benefit of NDN router-side caching, since content is only intended to be received 
by one end-point.\footnote{However, caching remains useful in the event of packet loss.}

We observe that NDN, in its current state, is not well-suited for this type of communication.
Even though, as the above description shows, bi-directional communication can be implemented
in NDN, the result is somewhat awkward and inefficient. Consider what it takes for Alice and Bob
to exchange one content packet
in each direction: (1) Alice issues/forwards an interest in Bob's next content, 
(2) Bob replies with requested content, (3) Bob issues an interest in Alice's next content, and, finally,
(4) Alice replies. All this requires two interests and two content packets. Each of these packets
traverses a sequence of NDN routers and requires separate processing. For each router, 
an interest packet entails~\cite{ccnx-protocol}: \\

\fbox{\begin{minipage}{0.49\textwidth}
\begin{quote}\small\sf
(i.1) packet reception from layer below \\
(i.2) PIT look-up (existing entry for the same name?)\\
(i.3) creation of a new PIT entry \\
(i.4) FIB look-up, and \\
(i.5) forwarding to next hop
\end{quote}
\end{minipage}}\\
\vspace{0.2cm}

Note that (i.2) is designed to collapse duplicate interests, i.e., those issued by different consumers
for the same content. However, in our setting of a point-to-point session, 
(i.2) is nearly useless, since only Alice issues interests for Bob's content and vice-versa.
Its only remaining utility is in preventing inadvertent retransmissions.
Also, given that numerous interests flow from Alice to Bob as part of the same session, 
performing (i.4) for each interest seems wasteful.\footnote{Of course, routing could conceivably 
change while a session is in progress; we discuss this Section~\ref{sec:design-usingpiggyback}. }

Furthermore, for each content packet in either direction, every intervening router must perform: \\

\fbox{\begin{minipage}{0.48\textwidth}
\begin{quote}\small\sf
(c.1) packet reception from layer below \\
(c.2) PIT look-up: content name$\rightarrow$pending interest\\ 
(c.3) caching content in CS \\
(c.4) deletion of a PIT entry, and \\
(c.5) forwarding to next hop, i.e., where the original interest came from
\end{quote} 
\end{minipage}}\\
\vspace{0.2cm}

In summary, to exchange a content packet in each direction, each router must perform (i.1)-(i.5)
and (c.1)-(c.5) operations twice. We believe that the corresponding overall 
amount of ``work'' is excessive and can be optimized for better performance. 
More generally, we claim that making NDN more friendly to bi-directional point-to-point 
communication is worthwhile, since, as discussed earlier, applications that involve this
type of communication are here to stay, and their requirements are quite distinct from those
of content distribution. To this end, in the next section we present an NDN add-on 
technique that offers appreciably better performance for bi-directional point-to-point 
communication while retaining all current features of NDN. 

\section{Design}
\label{sec:design}
Our approach adds a new (third) packet type to the NDN architecture, 
which we refer to as a {\em piggyback packet}, geared specifically for 
conversation-style applications. A piggyback packet essentially bundles  
a content packet with an interest packet, both of which travel to the same
end-point. Intuitively, the main idea is as follows: 
\begin{quote}
Suppose that Bob receives an interest for its next content (e.g., keystroke or 
voice frame, depending on the application) from Alice. Suppose that Bob also needs 
to obtain Alice's next content. Then, instead of responding to Alice's interest with a 
content packet and separately issuing an interest for Alice's next content, Bob piggybacks 
the latter onto the former. 
\end{quote}
Using piggyback packets offers two main benefits:
\begin{enumerate}
\item Fewer packets means
that there are fewer invocations of the lower/higher-layer API-s, i.e., steps (i.1)/(c.1) and
(i.5)/(c.5) are conjoined
\item No FIB look-up (i.4) needs to be performed for a piggybacked interest 
since it travels along with content for which router PIT state already
indicates the next hop. 
\end{enumerate}
In more detail, processing a piggyback packet by an NDN router
involves the following actions: \\

\fbox{\begin{minipage}{0.48\textwidth}
\begin{quote}\small\sf
(p.1) packet reception from layer below \\
(p.2) PIT look-up: content name$\rightarrow$pending interest \\
(p.3) caching piggybacked content in CS \\ 
(p.4) create new PIT entry for piggybacked interest \\
(p.5) deletion of the original PIT entry, and \\
(p.6) forwarding piggyback packet to the next hop 
\end{quote}
\end{minipage}}\\
\vspace{0.2cm}

Compared with steps (c.1)-(c.5) and (i.1)-(i.5) to process a content and an 
interest packet separately, processing a piggyback packet appears to be
much more efficient. In fact, the only extra action performed by a router
over and above processing a separate content packet is (p.4) -- creation of a new
PIT entry while saving all the processing steps associated with an interest packet. 
At the same time, we note that sending a piggyback packet is functionally 
equivalent to sending the interest and content separately.
The inclusion of the piggyback packet leaves the original  NDN architecture
unchanged.  

The rest of this section describes our design choices for constructing and processing 
piggyback packets.

\subsection{Constructing Piggyback Packets}
\label{sec:design-encoding}
We consider two approaches to forming piggyback packets, referred to as
concatenation and embedding.

The most obvious way to form a piggyback packet is to simply concatenate an interest
packet to a content packet but leave the two essentially intact, 
except for a flag in the content header indicating that this is a piggyback packet.

The second approach is similar but the interest is now embedded within 
the content packet. In this case, additional fields are needed to specify the interest offset. 
The main functional distinction of this approach is that the entire packet (both content and interest parts)
are covered by the producer's (sender's) signature. This has certain security implications which we
discuss below.

There are some obvious trade-offs between concatenation and embedding. 
The former allows more flexibility since any router that processes a piggyback packet 
can easily decouple interest and content packets and forward them separately. 
This allows routers to apply local policy and choose whether to treat piggyback packets
as one unit or unbundle them. It also allows piggyback support to be introduced incrementally.
For example, an NDN router that knows that its next-hop neighbors does not support piggybacking 
must decouple piggyback packets before forwarding.

In case of embedding, this flexibility is lost since the producer's signature covers the
entire piggyback packet and decoupling is impossible without violating one of the main
NDN tenets -- verifiability of content packet signatures by any NDN entity, including routers.
(Clearly, a router can not recompute or modify the content producer's signature).
On the other hand, embedding offers better overall security since interests can be authenticated
along with content. This might help in mitigating so-called {\em interest flooding} attacks \cite{ndn-dos}.

Yet another potential consideration is privacy: NDN interests are not signed by design since a 
digital signature leaks its source, i.e., the signer's (content producer's) identity. In case of 
embedding, interests
are signed along with content. However, we observe that privacy of NDN interests is most
relevant in the context of distribution of popular content, i.e., situations where multiple
consumers request the same content. This is very different from our setting of point-to-point
bi-directional communication where both end-points are well aware of each other. 
We also consider the privacy issue from the perspective of NDN routers. Normally,
an NDN router does not learn the identity of the source of an interest (content consumer). 
It only learns the identity of the producer of content that satisfies that interest.
However, in the context of processing a piggyback packet,
an NDN router learns that the content and the piggybacked interest are originated by
the same entity. 

We also note that, in terms of privacy, there is almost no difference between concatenation 
and embedding. Regardless of whether or not a piggybacked interest is covered by a signature, 
any entity that sees a piggyback packet clearly learns the origin of both the content and the 
interest components.

While some loss of privacy seems to be inherent to the use of piggyback packets, 
there are also potential avenues for security and privacy improvements per directional flow. 
For example, routers can apply certain policies to protect conversational flows from 
eavesdroppers; they can in fact discard interests received off-path, thus preventing any 
additional parties from accessing session content. This might, however, impact 
error recovery mechanisms, where a route change could lead an end-point to re-issue 
an interest over a different path.

\subsection{Using Piggyback Packets}
\label{sec:design-usingpiggyback}
As mentioned earlier, applications that involve continuous bi-directional (point-to-point) 
communication are the ideal candidates for taking advantage of piggyback packets. Such 
applications generate interest and content packets in a synchronous manner, 
with data continuously flowing in both directions. Audio/video conferencing is 
one example of this application class. 
These applications tend to have strict timing constraints on bandwidth and latency in order to
ensure satisfactory end-user experience. As an illustration, we consider an audio 
conferencing application running over plain NDN -- without piggybacking. We then show 
how piggybacking helps and discuss some issues related to packet loss.
\begin{enumerate}
\item Alice and Bob each launch their audio-conferencing application, registering their respective 
namespace to receive interests and publish content (voice data): \verb|/ndn/com/abc/alice/voice| and 
\verb|/ndn/edu/xyz/bob/voice|.
\item Alice initiates a call to Bob by issuing an interest for: \verb|/ndn/edu/xyz/bob/voice/call/alice|.
\item Bob receives this interest, parses the suffix identifying Alice and accepts the call, 
responding with a content packet as an acknowledgment.
\item Bob sends an interest anticipating content from Alice. The first such interest is: 
\verb|/ndn/com/abc/alice/voice/call/bob/0|. The trailing component (``\verb|0|'') represents
the initial sequence number of Alice's content which Bob wants to retrieve. 
\item Alice responds with the initial content and also issues an interest expecting 
content from Bob: \verb|/ndn/edu/xyz/bob/voice/call/alice/0|.
\end{enumerate}
Each party generates content at some negotiated rate, e.g. every 20ms (i.e., $rate = $ 50 pkts$/$s).
Similarly, 
interests are issued with increasing sequence numbers to keep pace with available content. 
Typically a {\em sliding window} mechanism is used to achieve pipelining. 
The window size $w$ (corresponding to maximum number of outstanding interests), 
is selected such that the streaming of content overlaps with the round-trip time (RTT), i.e., 
$w = \lceil RTT\cdot rate\rceil$.  This ensures that both parties receive a continuous 
audio stream throughout the session. 

We note that, in Steps 3 and 4,  Bob can start taking advantage of piggyback packets 
by combining the acknowledgement and his first interest. For her part, upon receipt of Bob's 
first piggyback packet, Alice can respond with a piggyback packet with her own audio content 
and an interest requesting audio content from Bob. Hereafter, both parties continue to 
exchange piggyback packets in lock-step, until the end of the session. 

An important advantage of using piggyback packets is that each intervening NDN router 
performs {\bf only one FIB look-up} for the entire session.\footnote{With a 
sliding window, $w$ 
route lookups are performed.} This single FIB look-up is performed at 
Step 1, when Alice issues her initial interest to Bob. This represents substantial savings for 
NDN routers, resulting in  overall lower latency. Also, as a consequence of using 
piggyback packets in lock-step, Alice and Bob use the same route
both directions; this route is ``fixed'' by Alice's initial interest. (Albeit, if with a sliding window
of size $w$, up to $w$ routes might be used.) This provides a reliable measure for RTT
between Alice and Bob as determined by Alice upon receipt of the first piggyback packet.

The above scenario and expected savings occur under ideal network conditions, with 
stable routing and low congestion. Dynamic route changes and packet loss would 
certainly cause disruption. To recover  from such events, the application can (and should)
temporarily revert to sending interest and content packets separately. 
Piggybacking can be resumed once a new reliable path between the parties is established.

The above scenario shows that audio/video conferencing applications are naturally predisposed 
to the use of piggyback packets. However, we emphasize that this technique is equally applicable to 
any applications requiring low-latency bi-directional communication. For example, a file transfer 
application can combine interests for additional content with acknowledgement packets. 
Also, an interactive chat application can bundle keep-alive (content) and interest packets 
if there is no actual data ready to be sent to prevent PIT entries created by bundles from expiring.

\section{Implementation}
\label{sec:piggyback-implementation}
Our implementation of piggyback packet support is based on the open-source 
NDN prototype, CCNx \cite{CCNx}. CCNx was originally developed at the 
Palo Alto Research Center (PARC) and was later made available to the academic 
and research community. We briefly overview CCNx, highlighting details 
relevant to our proposal. We then discuss modifications needed for piggyback packets.

\subsection{Overview of CCNx}
The main components of CCNx are the software forwarder (\verb|ccnd|) and the CCN client library
(\verb|libccn|). Both are implemented in C.
Packets in CCNx have a free-form structure and allow for variable sized fields; the prototype
does not impose any limitation on field size or packet length. Packets are encoded using a 
compact binary XML representation known as CCNx Binary Encoding (``{ccnb}'') 
\cite{ccnb-specifications}, which is designed specifically for CCNx. The client library 
provides full support for encoding and decoding ccnb-structured data and provides 
high-level API functions to create ccnb-encoded interest and content packets.

The \verb|ccnd| forwarder implements the NDN router functionality, including FIB, 
PIT, and Content Store (CS). All packets are sent and received through a ``face''. A 
face extends the notion of a network interface to include virtual transport 
mechanisms such as TCP/UDP tunnels and inter-process communication 
mechanisms. In addition to interfacing with other routers, this allows applications 
to communicate directly through a face. The forwarder is agnostic as to whether an
application or another forwarder is on the other end of a given face.

Currently, \verb|ccnd| forwarder lacks support for sending packets directly over a 
physical network interface. For now, forwarders connect to each other using long-lived 
TCP/UDP sessions, forming an overlay network on top of IP. A utility to inject and delete 
FIB entries is provided. When a route is entered into a FIB, a corresponding face is 
created between forwarders for routing interest and content packets. Similarly, 
applications register a face to send and receive packets to/from the network. 
Applications must have a local forwarder instance running on the same host in order 
to communicate over the network.

Applications register namespaces, encode/decode interests and publish content using
the CCNx client library, that also supports directly connecting to the local 
\verb|ccnd| forwarder using an IPC socket.

\subsection{Implementing Piggyback Support}
In order to extend CCNx to support piggyback packets, we: (1) define a new packet structure; 
(2) define new encoding and parsing functions in CCNx client library; and (3) modify the 
forwarder to support the new packet type. Our
implementation\footnote{The modified version is publicly available at \url{http://sprout.ics.uci.edu/projects/ndn/resources.html}.}
 is based on CCNx version 0.5.1.

As discussed in Section \ref{sec:ndn-overview}, we consider two approaches -- concatenation
and embedding -- to forming a 
piggyback packet. Although they can coexist, we currently only provide support for concatenation.
The encode function takes a ccnb-encoded interest and content as input and 
returns a ccnb-encoded piggyback packet. The packet structure defines a 
distinguishing tag labelled \verb|Piggyback| to allow parsers to distinguish it
from \verb|Interest| and \verb|Content| tags. The rest of the layout is followed 
by two tagged arbitrary-size binary objects containing the corresponding 
content and interest ccnb-encoded data. Textual XML representation of a piggyback 
packet is as follows: \\

\begin{minipage}{0.448\textwidth}
\hrulefill
\begin{verbatim}
<Piggyback>
    <Content>
        [ccnb-encoded content data...]
    </Content>
    <Interest>
        [ccnb-encoded interest data...]
    </Interest>
</Piggyback>
\end{verbatim}
\hrulefill\\
\end{minipage}

A piggyback packet parser function is also included. It takes a piggyback packet and extracts 
encapsulated interest and content packets returning a copy of both as ccnb-encoded data.

An application sends a piggyback packet in the same manner as an interest. We therefore duplicate the 
functionality of issuing interests, and allow consumers to issue piggyback packets instead. If a piggyback 
packet is issued, the interest packet is extracted, registered in the client PIT along with a callback handler 
to process the corresponding incoming content. Then, the piggyback packet is delivered over the 
IPC socket connecting the forwarder.

The \verb|ccnd| forwarder runs a tight event loop which does the following: 
\begin{enumerate}
\item polls each face for incoming data and assembles ccnb-encoded packets, 
\item processes packets according to their type, and 
\item forwards any outstanding data.
\end{enumerate}
The code that handles (1) and (3) is agnostic with respect to the data being sent or received. 
Therefore, we only needed to modify (2) to include support for piggyback packets.

The packet processing function parses a ccnb-encoded data stream and determines
whether the opening tag is \verb|Interest| or \verb|Content|. It then calls the appropriate 
handler. We add a branch to distinguish a \verb|Piggyback| tag, and call a custom
piggyback handler function.

The piggyback handler extracts the encapsulated interest and content components. 
First, a modified interest handler function is called, duplicating the functionality of the 
original interest handler, except for propagating the interest. The forwarding look-up 
is retained in case the content does not satisfy any interests. The modified content 
handler is called with both the content and piggyback messages. The content is then
processed in the same manner as in the original handler, except the piggyback packet 
is cached in the CS. If an interest is satisfied, the piggyback packet is placed into the
outbound queue of the outgoing face.

\section{Experiments}
\label{sec:experiment}
Using our implementation as described in Section \ref{sec:piggyback-implementation}, we 
conducted various experiments to ascertain performance benefits of using piggyback packets. 
To do so, we built a prototype file exchange (bi-directional transfer) application between two users. 
This application comes in two flavors: with and without piggyback support. To perform a fair 
comparison, the prototype with piggyback support is run exclusively on the modified 
CCNx code base. Whereas, the other prototype 
runs on the original unmodified CCNx code. 
We refer to the two application instances running on the users' hosts as A (Alice) and B (Bob). 

\begin{figure}[htb]
\centering
\includegraphics[scale=0.65]{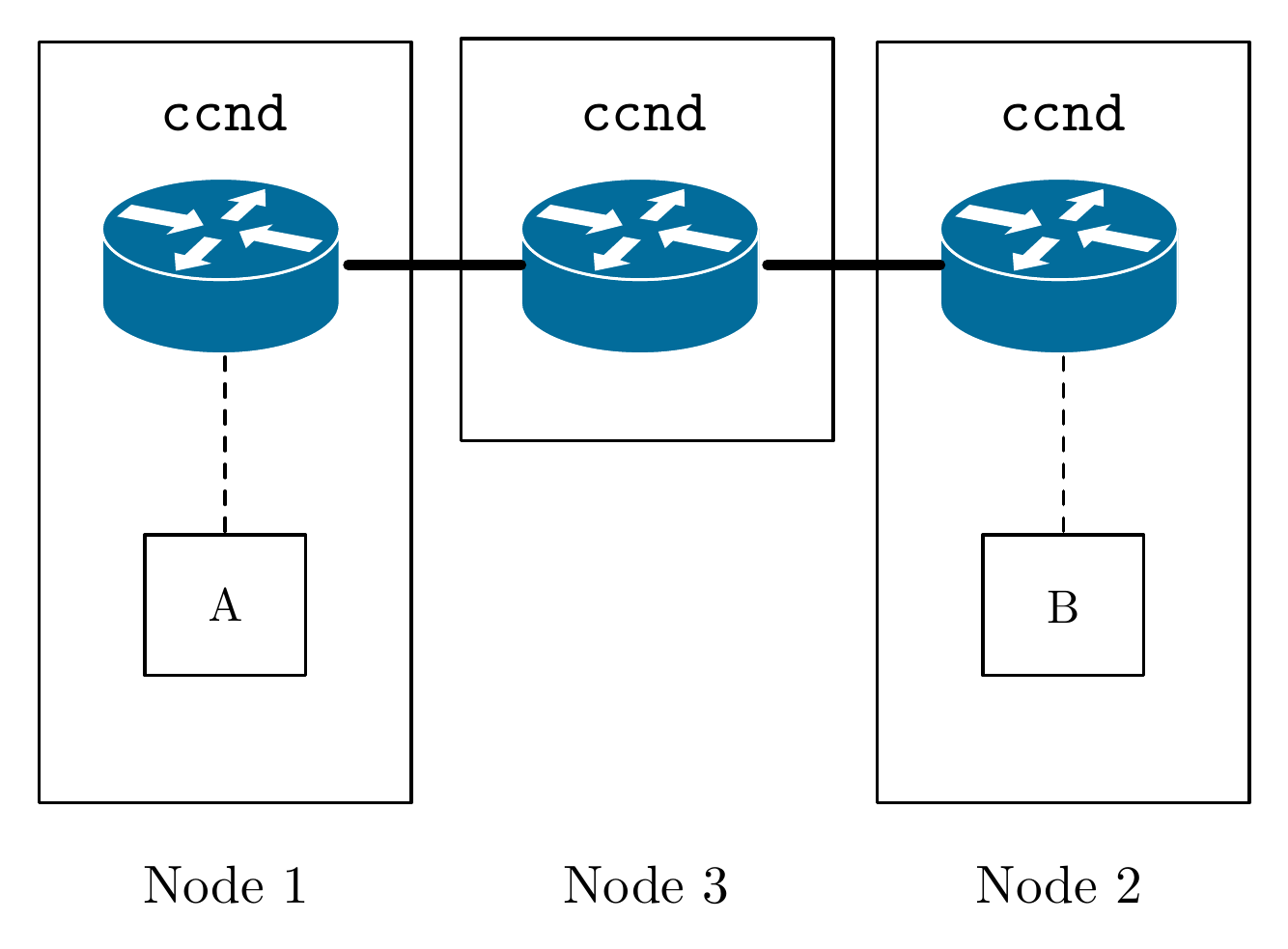}
\caption{Topology used in our experiments. Node 3 is the forwarder, Nodes 1 and 2 run 
applications A (Alice) and B (Bob), respectively.}
\label{fig:topology}
\end{figure}

Figure \ref{fig:topology} illustrates the topology used in our experiments. It is composed of three 
NDN nodes. The first node runs A and a copy of the \verb|ccnd| process that acts as a local NDN 
router. Similarly, the second node runs B and \verb|ccnd|. Finally, the third node forwards 
traffic from A and B  through its local copy of \verb|ccnd|.

Mutual file transfer is started by A, which sends an interest to B signaling the 
beginning of the process. B responds with the corresponding content and with its own 
interests for A's content.

The two parties send a total of 10,000 interests each, retrieving the same number of content packets. 
In our experiments, payload size of each content packet is 1,000 bytes. The total amount of data 
sent from A to B (and, from B to A) is about 10 MBytes.  We selected this size so that the 
size of all transferred content packets is below the MTU of our setup.  
We also performed additional experiments with content packets with different payload sizes, 
varying between 100 and 4,000 bytes. We omit these results due to space constraints, since they
are virtually identical to the one presented below.

Application A runs on a host with a quad-core Intel Xeon E5620 2.4GHz processors with 
12GB of memory. The forwarder (Node 3 in Figure~\ref{fig:topology}) runs on an machine 
with two quad-core Intel Xeon E5420 at 2.5GHz equipped with 16GB of RAM. 
Application B runs on a node equipped with an Intel Core2Duo 2.13GHz processor and 3GB 
memory. All machines run Ubuntu Linux and are connected using 100Mbps full-duplex 
Ethernet links with a maximum MTU of 1500bytes.

\subsection{Experiment Description and Performance Metrics}
\label{parameter}
As mentioned above, we performed experiments on bi-directional file transfer between two nodes. 
This setup retains all relevant properties of audio/video conferencing, while also
allowing us to vary the data rate. 

\begin{figure*}[htb]
\centering
\includegraphics[width=0.63\textwidth]{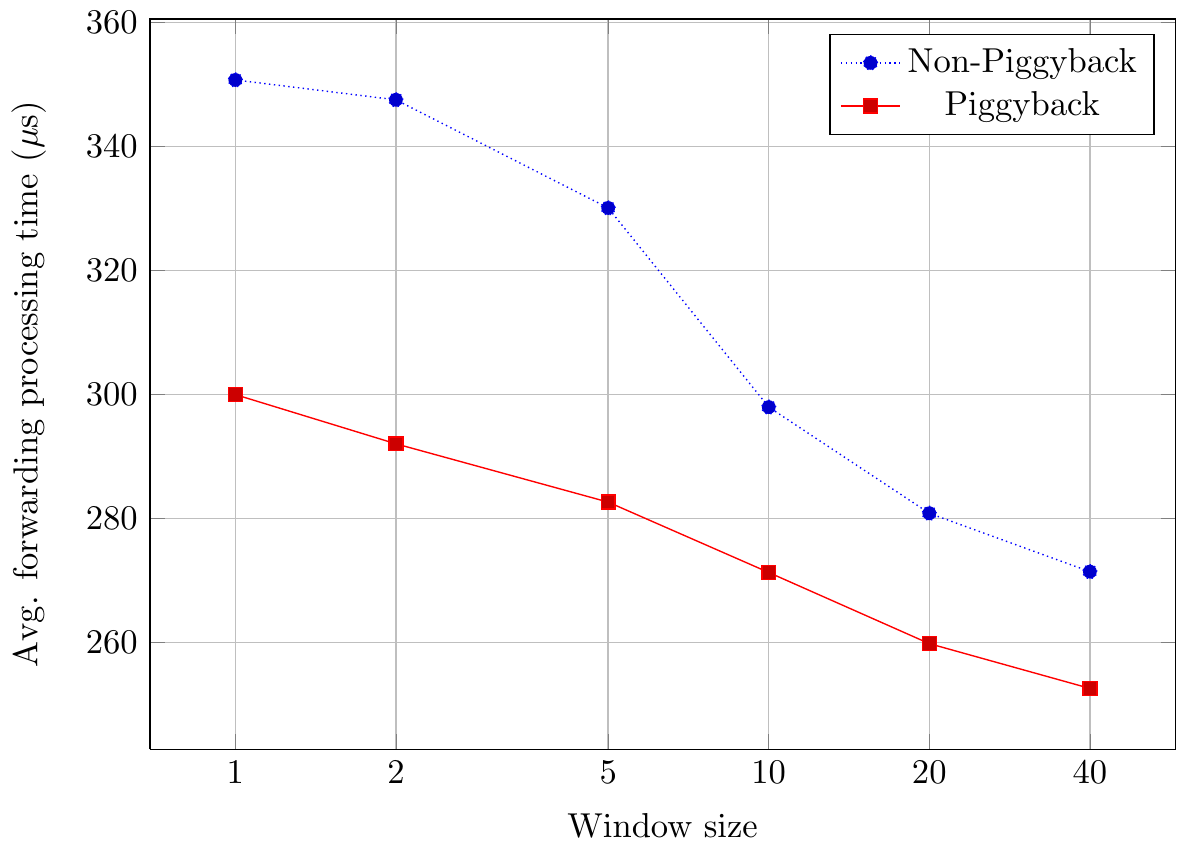}
\caption{Forwarding Processing Time (FPT) for various window sizes}
\label{fig:processing-1000}
\end{figure*}

In order to achieve maximum throughput our prototype uses pipelining, via a sliding 
window, as described in~\ref{sec:design-usingpiggyback}. In particular, for a windows 
size $w$, A sends $w$ interest packets to B. For every interest from A, B replies with 
the appropriate content packet followed by an interest for A's content. Similarly, for each interest 
received from B, A replies with a content packet and a new interest.  After that, A and B 
continue the exchange in lock-step until the entire transfer completes.

\begin{figure*}[htb]
\vspace{0.2cm}
\centering
\includegraphics[width=0.63\textwidth]{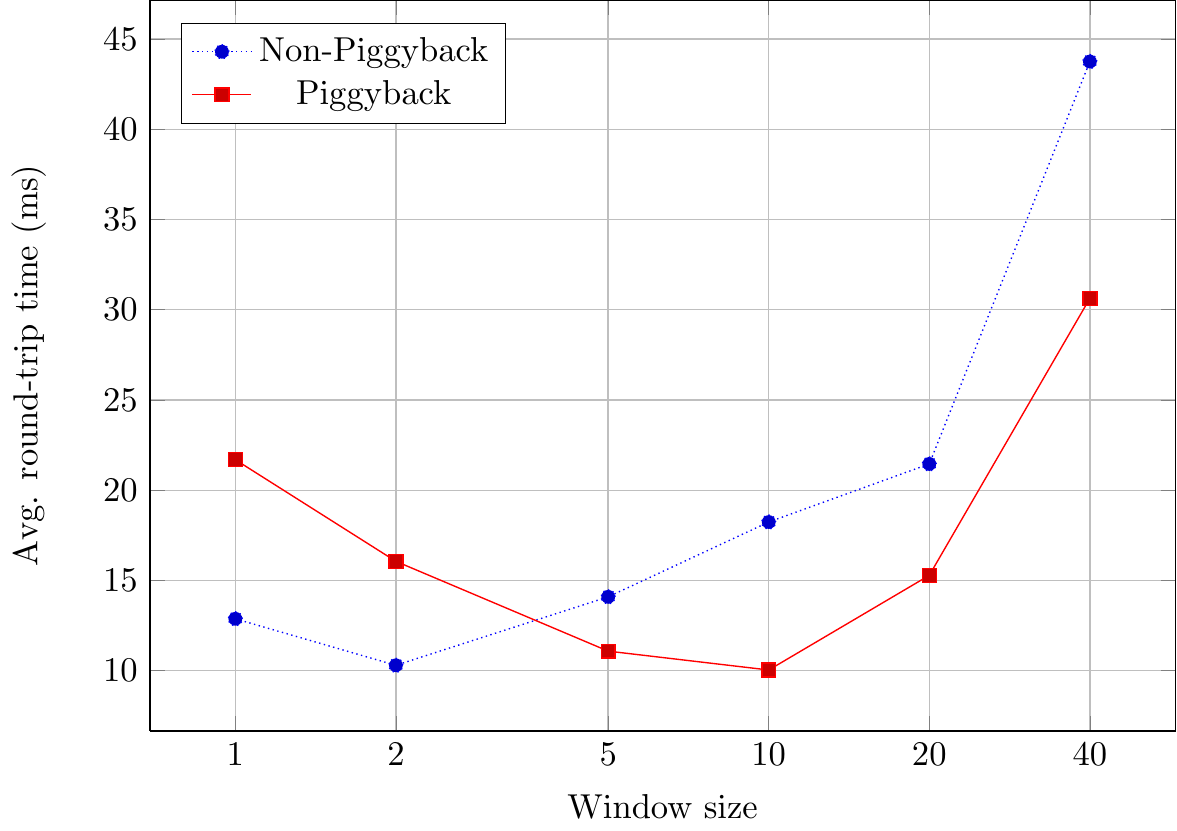}
\caption{Round Trip Time (RTT) for various window sizes}
\label{fig:rtt-1000}
\end{figure*}

When using piggyback packets and a window size of $w$, the very first $w$ interest packets from 
A and the very last $w$ content packets from B are not piggybacked. (In other words, they are sent
as interest and content packets, respectively). All other packets are piggybacks.
For example, if $w=1$, A sends one interest to B, then B responds with a piggyback packet. 
A issues its next piggyback packet after receiving the one from B. 
This goes on until the very last piggyback from A to B. Upon its receipt, B has no further interests 
to issue, so it sends a plain content packet to A.

Clearly $w$ is expected to affect packets processing  by the forwarder. A large  
$w$ might cause several packets to be queued in the forwarder's buffer, introducing delay. 
Since the choice of $w$ significantly affects performance, we run numerous tests,
varying $w$ between 1 and 40.
For each experiment, we measured the following: 
\begin{enumerate}
\item {\bf Forwarding Processing Time (FPT)}, which corresponds to the time for
Node 3 in Figure~\ref{fig:topology} to (logically) forward a content packet {\em and} and an interest packet 
in either direction. With piggybacking, we measured FPT as the time to forward 
a single piggyback packet. Without it,  FPT is measured as the sum of the times required to 
separately forward a content packet and an interest packet.

\item {\bf Round Trip Time (RTT)} --  time it takes for an interest to retrieve the corresponding 
content packet. 
When piggyback packets, we measured RTT as the time for a 
piggyback packet to retrieve the corresponding piggyback from the other side.
We note that this slightly penalizes the piggyback approach, since our measured RTT (in 
the piggyback case) corresponds to the transfer of a larger amount of data 
than in the non-piggyback case.

\item {\bf Transfer Time (TT)}, -- total time required to transfer all data involved in the 
experiments, i.e., 10MB of content from A to B.
\end{enumerate}

\subsection{Experimental Results}
\label{results}

\begin{figure*}[tb]
\centering
\includegraphics[width=0.63\textwidth]{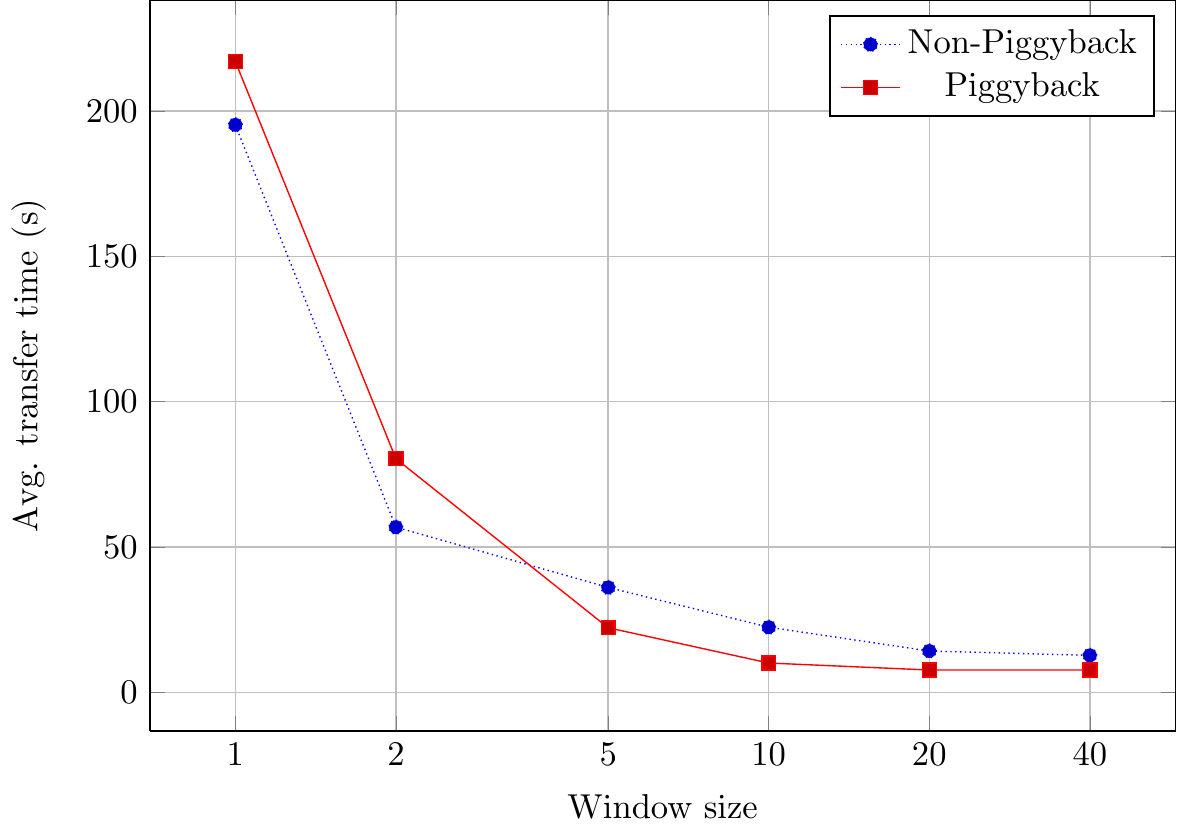}
\caption{Transfer Time (TT) for various window sizes}
\label{fig:throughput-1000}
\end{figure*}

We now report on the results of our experiments. 
Figure \ref{fig:processing-1000} illustrates FPT for both cases (with/without piggyback) 
for variable window sizes. Results show that, for all window sizes examined, FPT 
of a piggyback packet is smaller than that of an interest plus a content packet. In particular, 
for $w=1$, average processing time decreases by 14.5\% for the piggyback case. Meanwhile,
for $w=2$, piggybacking provides the biggest improvement as processing time decreases 
by 16\%. For larger window sizes, the difference between the two approaches becomes 
smaller; however, it remains significant with a minimum of 6.9\% for $w=40$.
This confirms that, by limiting layer 2-3 and API calls overhead, our piggyback solution allows forwarder to achieve significantly better performance.

Note that there is a decreasing trend in FPT as we increase window size for 
both cases. This is partially attributable to the way CCNx handles packets in 
its incoming buffer: as we increase window size, the number of invocations of 
packet (interest or content) processing routines in \verb|ccnd| decreases. 
In other words, when multiple packets are (temporarily) stored in a router's 
incoming buffer, \verb|ccnd| can pull and process many of them at once.

Figure \ref{fig:rtt-1000} shows average RTT for our experiments. As it illustrates, 
for $w<5$, RTT of non-piggybacked packets is lower. The reason for this is that 
piggybacked packets are larger than single interest or content packets.
In the non-piggybacked case, the smallest RTT is achieved when $w=2$.  This reflects our 
observation above: when \verb|ccnd|'s incoming buffer contains multiple packets, such 
packets are processed together, thus saving forwarding time. When $w=1$, the forwarder's 
buffer always contains at most one packet; with $w=2$, the buffer often contains two packets. 
However, we observe that further increasing 
$w$ does not provide additional benefits for non-piggybacked case, since savings in processing 
time are outweighed by the waiting time of multiple packets in the forwarder's buffer.

Since piggyback packets have lower processing overhead compared to 
interest/content packet pairs, the smallest RTT occurs
when $w=10$. This RTT is almost identical to that of non-piggybacked case 
with $w=2$. As shown below, larger window size implies better throughput. Therefore 
piggybacking allows our test application to achieve higher throughput with minimum latency.

Figure \ref{fig:throughput-1000} shows the transfer time (TT) for both piggyback and 
non-piggyback cases for various window sizes. For small $w$, TT  is 
higher for piggyback packets. This accounts for the fact that, in our setup, the 
optimal window size is greater than 2: in this case, RTT becomes a limiting factor for the
transfer rate, since the end-points end up waiting rather than issuing new interest 
and content packets. Therefore, smaller RTT guarantees better bandwidth 
for small $w$. However, increasing $w$ removes this effect. In particular, we 
observe that, for $w\geq5$, bandwidth available to the applications using piggyback 
packets is markedly higher than without piggybacking.

As expected, we observe that TT decreases for larger values $w$ for 
cases. In particular, for $w\geq5$ the transfer rate with 
piggyback packets is higher.

\section{Related Work}
\label{sec:related-work}
To the best of our knowledge, there has not been any previous work that addressed
combining requests and responses in bi-directional communication over CCN.

Somewhat related work has been done for TCP with selective acknowledgments 
(SACK)~\cite{rfc2018} and its extension, duplicate-SACK (D-SACK)~\cite{rfc2883}. 
SACK and D-SACK allow the receiver to selectively acknowledge correctly received packets, such
that the sender must only re-send the packets that have been lost. This is a significant improvement 
over older cumulative TCP acknowledgement techniques in presence of multiple packet loss from 
one window of data.

In \cite{clark91}, Clark et al.~study the effect of bi-directional traffic on TCP congestion control algorithm. In particular, they observe that in the BSD Tahoe TCP implementation packets from a single connection are clustered together, similarly to what observed by the same authors in~\cite{clark90} over one-way traffic. This causes ACK compression, which significantly reduces the available bandwidth for TCP connections. They also show that in case of two-way traffic, the issue of ACK compression is made worse by the interaction of ACKs and data packets in the queue.

In \cite{voccn} Jacobson et al.~address low-latency bi-directional traffic over NDN in the context of VoCCN, a real-time, conversational, telephony application over Content-Centric Networking (CCN). The authors show that NDN is capable of transporting this kind of traffic with jitter comparable to that of RTP. Although each packet is independently signed, voice data is timely delivered. Also, they argue that the content-centric model fits telephony applications better than the IP model.

\section{Conclusion and Future Directions}
\label{sec:conclusion}
In this paper we show that NDN, in its current state, is not well-suited for 
bi-directional low-latency point-to-point communication.  To provide better
efficiency for applications requiring this type of communication, we propose
some simple modifications. In particular, we introduce a new piggyback
packet type that bundles content with an interest traveling in the same direction. 
This results is roughly half the number of overall packets,  reducing the processing time in NDN routers,
which is reflected as an improved end-to-end throughput and reduced round-trip time.
Furthermore, the introduction of piggyback packets 
preserves the existing NDN architecture and is exposed to applications as an optional feature. We implement the proposed modification 
using CCNx -- a reference implementation of NDN. We also conduct extensive experiments 
that demonstrate
clear performance gains derived from using piggyback packets.

Clearly, this work is only intended as a first step towards efficient and dependable bi-directional point-to-point communication over NDN. Items for future work include the following:
\begin{itemize}

\item Performing additional experiments on more complex topologies, including the official NDN testbed~\cite{ndn-testbed}, to determine the impact of reduced processing time over longer routes.

\item Extending our model and implementation to handle bi-directional multicast traffic, such as audio/video conferencing among multiple users.

\item Evaluating the impact of many concurrent flows (rather than just a single one) between two or more nodes.

\item Measuring the effect of links with asynchronous upload/download bandwidth on piggybacking; determine how this impacts communication between hosts when the two parties need to exchange uneven amount of data.

\item Designing techniques that provide more sophisticated handling of packet loss, and  evaluating their  performance.

\item Exploring security and privacy trade-offs determined by the use of piggybacking.

\end{itemize}
Furthermore, we plan to identify (and experiment with) 
other classes of traffic that, similarly to bi-directional scenarios, 
can benefit from relatively small changes to the NDN architecture.

\bibliographystyle{IEEEtran}
\bibliography{references}

\end{document}